\documentclass[%
 reprint,
 amsmath,amssymb,
 aps,
]{revtex4-1}
\usepackage{subfig}
\usepackage{graphicx}
\usepackage{dcolumn}
\usepackage{bm}

\begin{document}

\preprint{APS/123-QED}

\title{Lattice Boltzmann simulations of apparent slip and contact angle in hydrophobic micro-channels}

\author{Renliang Zhang}
\author{Qinfeng Di}
 \email[Author to whom correspondence should be addressed. tel:(+86)-21-56333256; Electronic address: ]{qinfengd@sina.com}
\author{Guohua Gao}
 \altaffiliation{Also works for Shell as a senior petroleum engineer. His research interests include production optimization, automatic history matching, reservoir management and monitoring, and mechanics of tubulars.}
\author{Xinliang Wang}
\author{Weipeng Ding}
\author{Wei Gong}
\affiliation{%
 Shanghai Institute of Applied Mathematics and Mechanics, Shanghai University, Shanghai 200072, China\\
 Shanghai Key Laboratory of Mechanics in Energy Engineering, Shanghai 200072, China
}%

\date{\today}

\begin{abstract}
In this paper, we applied the Shan-Chen multiphase Lattice Boltzmann method to simulate two different parameters, contact angle (a static parameter) and slip length (a dynamic parameter), and we proposed a relationship between them by fitting those numerical simulation results. By changing the values of the strength of interaction between fluid particles ($G$) and the strength of interaction between fluid and solid surface ($G_{ads}$), we simulated a series of contact angles and slip lengths. Our numerical simulation results show that both $G$  and $G_{ads}$ have little effects on the relationship between contact angle and slip length. Using the proposed relationship between slip length and contact angle, we further derived an equation to determine the upper limit of nano-particles' diameter under which drag-reduction can be achieved when using nano-particles adsorbing method.
\end{abstract}

\maketitle


\section{\label{sec:level1}Introduction}

Contact angle is a static parameter of measuring the wettability of a liquid on a solid surface, and it can be easily measured. Slip length is a dynamic parameter of quantifying the non-zero velocity boundary condition of a liquid flowing over a solid surface. Determination of slip length is very important for calculation of drag and other hydrodynamic behaviors of fluid flowing through micro-channels or over nano-scale patterned surfaces. However, it is very difficult to directly measure the apparent slip length.

Non-slip boundary condition is extremely successful in describing macro-scale viscous flows in engineering applications for more than one hundred years\cite{kunert2008effect}.This assumption is supported only by macroscopic experimental results. However, a series of experiments\cite{qinfeng2009experimental,xinliang,burton2005hydrophobicity} and numerical simulation results\cite{szalm¨¢s2006slip,
zhang2010lattice,voronov2007slip} indicate that this assumption does not hold at micro- and nano-scale, and a slip boundary condition should be applied. Based on this slip boundary effect, artificial super-hydrophobic surfaces have been widely used in industrial production and people's daily lives, for example, self-cleaning paints, roof tiles, fabrics and glass windows that can be cleaned by a simple rainfall\cite{lai2003mimicking} and the nano-particles adsorbing method\cite{qinfeng2009experimental} in improving oil recovery are all in practice.

In 1823, Navier proposed a slip boundary condition that the fluid velocity at a point on a surface is proportional to the shear rate at the same point,$v(x_b)=\delta\partial v(x)/\partial x$, where $\delta$ is defined as the slip length, but its value is hard to determine. Molecular dynamics simulations (MDS) have been widely used to study the relationship between fluid slip and the properties of fluid and solid, and the results proved that there is boundary slip on microscopic scale\cite{cottin2003low}. However, the conventional MDS method has difficulty to determine the small flow velocity because the nonlinear coupling of the small bulk flow velocity with the large peculiar velocity of the thermal motion\cite{zhang2008low} and has difficulty to simulate large size systems \cite{sbragaglia2006surface}. Chen et al investigated the Couette geometry flows by means of a two-phase mesoscopic Lattice Boltzmann(LB) model, and the results show that there is a strong relationship between the magnitude of slip and the solid-fluid interaction\cite{yan2008boundary}, but it is a very difficult or even an impossible task to compute the exact slip in dependence of interaction and it is also difficult to apply to engineering.

In this present work, we focus on investigating the effects of wall wettabilities on the slip length. With a general bounce-back no-slip boundary condition applied to the interface between fluids and solid surfaces, together with the Shan-Chen multiphase model\cite{shan1994simulation}, the LB method is used to simulate the Poiseuille flow. The simulation results reveal that the wetting properties of the wall have an important effect on the flow. We proposed a relationship between slip length and contact angle according to these numerical simulation results. Using this relationship, it is easy to estimate the slip length because the contact angle is a parameter that can be easily measured.

Nano-partical adsorption has been proved an effective drag-reduction method that can be widely used for enhancing oil recovery and many other practical applications. However, it is still not clear how to properly select the size or diameter of nano-particles. The final section considers drag reduction by hydrophobic nanoparticles (HNPs). Placing these particles on the walls of a channel constricts the channel but increases the slip length. An overall reduction in flow resistance occurs if the latter effect outweighs the former. Using the relationship between slip length and contact angle, we further derived an equation to determine the upper limit of nano-particles' diameter under which drag-reduction can be achieved.
\section{Numerical model}
\subsection{\label{sec:level2}The LBGK model}
The LB method, which involves a single relaxation time in the Bhatnagar-Gross-Krook (BGK) collision operator\cite{qian1992lattice}, is used here. The time evolution of this model can be written as
\begin{equation}
f_i({\bf x}+{\bf c}_i\Delta t,t+\Delta t)-f_i({\bf x},t)=-\frac{1}{\tau}[f_i({\bf x},t)-f_i^{eq}({\bf x},t)],
\end{equation}
where $f_i({\bf x},t)$ is the single-particle distribution function for fluid particles that move in the direction ${\bf c}_i$ at $({\bf x},t)$, $f_i^{eq}({\bf x},t)$ is the equilibrium distribution function, $\Delta t$ is time step of simulation, and parameter $\tau$ is the relaxation time characterizing the collision processes by which the distribution functions relax towards their equilibrium distributions.
In the two-Dimensional (2D) nine-particle model, the equilibrium distribution function, $f_i^{eq}({\bf x},t)$, depends only on local density and velocity and can be chosen as the following form
\begin{equation}
f_i^{eq}=\omega_i\rho \left [ 1+\frac{({\bf c}_i\bf{\cdot u})}{c_s^2}+\frac{({\bf c}_i\bf{\cdot u})^2}{2c_s^4}+\frac{(\bf {u\cdot u})}{2c_s^2}\right ],
\end{equation}
where
\begin{eqnarray}
{\bf c}_i=\left \{
                          \begin{array}{cc}(0,0)c & i=0 \cr
                          [cos\frac{(i-1)\pi}{2}],sin\frac{(i-1)\pi}{2}]c & i=1,2,3,4 \cr
                          \sqrt{2}[cos\frac{(2i-9)\pi}{4}],sin\frac{(2i-9)\pi}{4}]c & i=5,6,7,8
                          \end{array}
          \right.,
\end{eqnarray}
\begin{eqnarray}
\omega_i=\left \{
                          \begin{array}{cc}4/9 & i=0 \cr
                          1/9 & i=1,2,3,4 \cr
                          1/36 & i=5,6,7,8
                          \end{array}
          \right.,
\end{eqnarray}
\begin{eqnarray}
c_s^2=\frac{1}{3}c^2,
\end{eqnarray}
$c=\Delta x/\Delta t$ is lattice velocity, $\Delta x$ is lattice distance, and $\Delta t$ is time step of simulation.

The mass density $\rho$ and momentum density $\rho \bf u$ of the fluid are calculated from the first and second moments of the distribution function, i.e.,
\begin{eqnarray}
\rho=\sum_i f_i,
\end{eqnarray}
\begin{eqnarray}
\rho {\bf u} = \sum_i {\bf c}_if_i,
\end{eqnarray}
And the relaxation time tunes the kinematic viscosity as
\begin{equation}
\nu=(\tau-\frac{1}{2})c_s^2\Delta t,
\end{equation}
The non-slip boundary condition at solid-fluid interfaces is realized by the bounce-back rule, where the momentum of the fluid particles that meet a solid wall is simply reversed. The bounce-back rule is simple and computationally efficient, and enables fluid flow simulations in complex geometries. On the inlet and outlet boundaries, the periodic boundary conditions are used.

\subsection{\label{sec:citeref}Shan-Chen multiphase model}
To simulate non-ideal multiphase fluids, the attractive or repulsive interaction among molecules, which is referred to as the non-ideal interaction, should be included in the LB model. There are many approaches to incorporate non-ideal interactions, such as color-fluid model, interparticle-potential model, free-energy model, mean-field theory model and so on. The interparticle-potential model proposed by Shan \& Chen\cite{shan1994simulation,shan1993lattice} is to mimic microscopic interaction forces between the fluid components. This model modified the collision operator by using an equilibrium velocity that includes an interactive force. This force guarantees phase separation and introduces surface-tension effects\cite{aidun2010lattice}.

This model has been applied with considerable success in measuring contact angles\cite{huang2007proposed} and the effect of wall wettabilities, topography and micro-structure on drag reduction of fluid flow through micro-channels\cite{zhang2010numerical}. As an extension of the Shan-Chen model, Benzi\cite{benzi2006mesoscopic} first derived an analytical expression for the contact angle and the surface energy between any two of the liquid, solid and vapor phases.

In the Shan-Chen multiphase model, the non-ideal interaction is obtained by using an attractive short-range force
\begin{equation}
{\bf F}({\bf x})=-G\psi({\bf x},t)\sum_i w_i\psi({\bf x}+{\bf c}_i\Delta t,t){\bf c}_i,\label{209}
\end{equation}
Adhesive forces between the fluid and solid phases are modeled by introducing an extra force
\begin{equation}
{\bf F}_ads({\bf x})=-G_{ads}\psi({\bf x},t)\sum_i w_is({\bf x}+{\bf c}_i\Delta t,t){\bf c}_i,\label{210}
\end{equation}
where
\begin{equation}
w_i=\left \{
                          \begin{array}{cc}0 & i=0 \cr
                          1/9 & i=1,2,3,4 \cr
                          1/36 & i=5,6,7,8
                          \end{array}
          \right.,
\end{equation}
Here  $s({\bf x},t)=0,1$ for the fluid and the solid phase, respectively, $G$, the interaction strength, is used to control the two-phase liquid interaction, and the adhesion parameter $G_{ads}$ is used to control the wettability behavior of the liquid at solid surfaces. $\psi({\bf x},t)$ is a local 'effective mass'\cite{shan1994simulation,shan1993lattice,qian1995recent,sukop2006lattice}, which is defined as\cite{sukop2006lattice}
\begin{equation}
\psi({\bf x},t)=\psi_0 e^{-\rho_0/ \rho},
\end{equation}
Using these definitions, the fluid momentum is changed at each time step according to
\begin{equation}
\rho {\bf u}^{\textquoteright}=\sum_i{\bf c}_if_i+\tau({\bf F}+{\bf F}_{ads}),
\end{equation}
where ${\bf u}^{\textquoteright}$ is the new fluid velocity. The equation of state in the Shan-Chen model is\cite{sukop2006lattice}
\begin{equation}
P=\rho RT+\frac{GRT}{2}[\psi(\rho)]^2,
\end{equation}

\section{Numerical simulation of contact angle}
\begin{figure}[h]
\includegraphics[scale=.6]{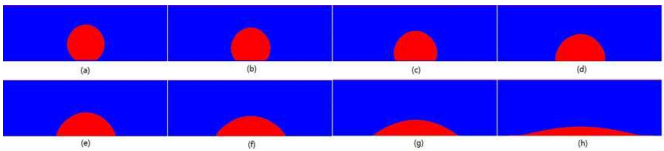}
\caption{\label{fig1} Static contact angle.}
\end{figure}

The LB simulations were carried out in a 2D domain.The grid mesh used is $50\times200$. In the simulation,the general non-slip bounce-back scheme was employed for the solid-fluid interfaces, and periodic boundary conditions were applied at both inlet end and outlet end along the horizontal direction. A droplet, whose diameter is 30, is set on the middle between two ends.After 30000 time steps (in our simulations the units are lattice units as Ref.~\onlinecite{sukop2006lattice}, and the same as below), the result tends to stabilize. FIG.~\ref{fig1} shows different static contact angles for $\psi({\bf x},t)=4e^{-200/ \rho}$\cite{sukop2006lattice}, in Eqs.~(\ref{209}) and~(\ref{210}). Values of parameters $G$ and $G_{ads}$ and the simulated contact angles (in degree) for each case are listed in TABLE~\ref{tab1}.

\begin{table}[h]
\caption{\label{tab1}%
 Adhesion parameters $G_{ads}$ and the contact angles (in degree) for droplets.
}
\centering
\begin{ruledtabular}
\begin{tabular}{p{30pt}p{50pt}p{65pt}p{65pt}}

Case & Adhesion parameters & Contact angle for $G=-120$ & Contact angle for $G=-130$
\\
\colrule
a & -100 & 147.2 & 147.8\\
b & -130 & 127.6 & 132.4\\
c & -160 & 109.8 & 117.6\\
d & -190 &  92.6 & 102.6\\
e & -220 &  75.3 &  87.6\\
f & -250 &  58.6 &  73.1\\
g & -280 &  40.7 &  59.5\\
h & -310 &  18.0 &  45.3\\
\end{tabular}
\end{ruledtabular}
\end{table}

As shown in FIG.~\ref{fig1}, $G$ and $G_{ads}$ determine the value of contact angle. $G_{ads}$ represents the strength of interaction between fluid and solid surface and $G$ represents the strength of interaction between fluid particles. A negative $G_{ads}$ indicates attractive interaction. When $G$ is given, the greater the magnitude of $|G_{ads}|$, the stronger the reaction, and thus resulting in smaller contact angle. So we can change parameter $G$ and parameter $G_{ads}$ to simulate contact angle for arbitrary surface condition, and then obtain different wall wettabilities. Form TABLE~\ref{tab1}, we see that both parameters of $G$ and $G_{ads}$ have significant impact on the simulated contact angle.

\section{Numerical simulation of slip length}
\begin{figure}[h]
\includegraphics[scale=.35]{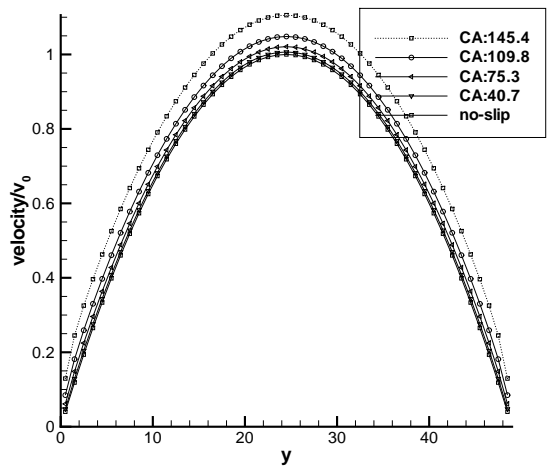}
\caption{\label{fig2} Velocity profiles with different contact angles.}
\end{figure}

\begin{figure}[h]
\includegraphics[scale=.35]{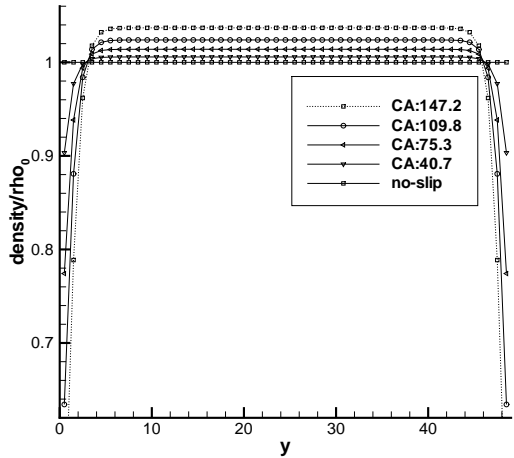}
\caption{\label{fig3} Density profiles with different contact angles.}
\end{figure}

In order to investigate the slip effect of boundary, we conducted numerical simulations of 2D Poiseuille flows. Typical density and velocity profiles of pressure-driven Poiseuille flows are displayed in FIG.~\ref{fig2} and FIG.~\ref{fig3} (given $G=-120$). The horizontal coordinate in both FIGs. represents the distance measured from one of the solid surface boundary to the other. The vertical coordinate in FIG.~\ref{fig2} is the normalized velocity, where v$_0$ is the max velocity measured at the center in the channel for the case of no slip. The vertical coordinate in FIG.~\ref{fig3} is the normalized fluid density, where rho$_0$ is the density of liquid for the case of no slip. The pressure gradient is specified as 0.005 in lattice unit for both cases. Different contact angles (as shown in both FIGs~\ref{fig2} and~\ref{fig3}) can be simulated by specifying different values of the adhesion parameter ($G_{ads}$).All simulations were run until static equilibrium was nearly attained, and then a pressure gradient of 0.005 was applied in the x-direction (flow direction).

As shown in FIG.~\ref{fig2}, fluid velocity approaches zero as $y\to 0$ (the lower boundary) or  $y\to 49$ (the upper boundary), which is consistent with the bounce-back boundary condition specified at the two boundaries. However, the fluid velocity increases dramatically in a very thin layer near the boundary. The layer is so thin that it is hard to see such details near the boundary, and velocity at the boundary looks like non-zero when plotted in a larger scale, as shown by plots in FIG.~\ref{fig2}. In a micro-scale, velocity at a boundary is zero, but in a macro-scale, the velocity appears non-zero (so called apparent slip). Plots shown in FIG.~\ref{fig2} clearly indicate that the slip velocity at the boundary increases as contact angle increases. In order to understand the physical mechanism of such kind of phenomenon, we also drew density profiles of fluid with different contact angles in FIG.~\ref{fig3}. We should note that the density of the fluid with zero contact angle is constant (as shown by the horizontal line in FIG.~\ref{fig3}). However, a sharp reduction of fluid density near the boundary is observed for a fluid with non-zero contact angle, which clearly indicates a layer of much less dense fluid (most probably gas) is induced between the dense liquid and solid surface. As discussed above, the parameter of $G_{ads}$ controls the interaction between the fluid and solid surface. Increasing $G_{ads}$  decreases the attraction (or increases the repulsion) between the liquid and solid surface, and thus attracts more gas to the surface. The less the dense of the fluid at the surface, the less viscous shear force, and the more significant slippage.

From FIG.~\ref{fig2} and FIG.~\ref{fig3} we can see that the wetting properties of the wall have an important influence on the velocity profile, especially the slip velocity at the boundary. The more hydrophobic the wall is, the larger the slip velocity is on the boundary. There is a low-density layer between the bulk liquid and the wall, and the more hydrophobic the wall is, the lower density of fluid is, see FIG.~\ref{fig3}. This result is similar to those obtained from other LB model simulation\cite{zhang2004apparent} and observed in MDS\cite{nijmeijer1990wetting}. Compared with the macroscopic flow, the ratio of the low-density layer region to the inner region is larger in the microscopic flow, and this is the main difference between micro-flow and macro-flow. Thus, the slip cannot be ignored in the micro-flow.

Following Navier's hypothesis, the velocity in flow direction at position $y$ between two parallel planes is given[1] by
\begin{equation}
v(y)=\frac{1}{2\mu}\frac{\partial P}{\partial x}(h^2-y^2-2h\delta),\label{eq}
\end{equation}
where $2h$ is the distance between the planes, $\mu$ is the viscosity, $\partial P/\partial x$ is pressure gradient, and $\delta$ is slip length. Both slip velocity ($v_0$) and slip length ($\delta$) can be estimated by matching the numerical LM simulation results of velocity profile with Eq.~(\ref{eq}).

\begin{figure}[h]
\includegraphics[scale=.5]{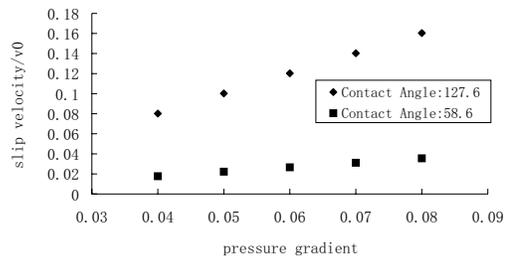}
\caption{\label{fig4} The slip velocity against pressure gradient.}
\end{figure}

\begin{figure}[h]
\includegraphics[scale=.5]{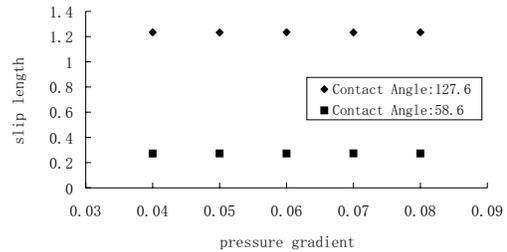}
\caption{\label{fig5} The slip length against pressure gradient.}
\end{figure}

FIG.~\ref{fig4} shows that slip velocity increases linearly as pressure gradient increases. The larger the contact angle is, the larger the slip velocity is achieved, as shown by the solid diamonds (a contact angle of 127.6$^\circ$ obtained by using  $G=-120$ and $G_{ads}=-130$) and squares (a contact angle of 58.6$^\circ$ obtained by using  $G=-120$ and $G_{ads}=-250$). However, the results shown in FIG.~\ref{fig5} clearly indicate that the estimated slip length is independent of pressure gradient.
\section{The relationship between slip length and contact angle}
\begin{figure}[h]
\centering
\includegraphics[height=4cm,width=8cm]{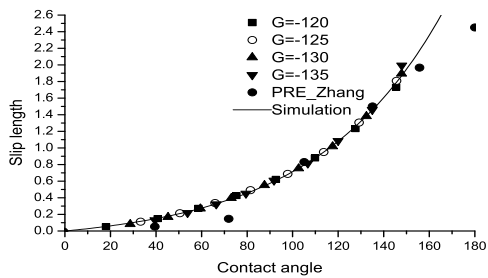}
\caption{\label{fig6}  The slip length against contact angle.}
\end{figure}

As discussed above, given an interaction strength ($G$), different contact angles and slip lengths can be simulated by using a series values of the adhesion parameter ($G_{ads}$) (see the values specified in TABLE~\ref{tab1}). Numerical results of contact angles and slip lengths, represented by different symbols with different colors shown in FIG.~\ref{fig6}, are corresponding to different values of $G$, respectively, squares ($G=-120$), triangles ($G=-125$), triangles ($G=-130$) and triangles ($G=-135$). The numerical results shown in FIG.~\ref{fig6} cover a wide range of contact angles, ranging from 18$^\circ$ to 150$^\circ$. Though completely different interaction strength between fluid particles ($G$) and different interaction strength between fluid and solid ($G_{ads}$) were used to simulate both contact angle and slip length, our numerical results (as shown in FIG.~\ref{fig6}) clearly indicate that slip length is a function of contact angle. The relationship between them can be expressed as Eq.~(\ref{eq1}) (as shown by the solid curve in FIG.~\ref{fig6}), which is obtained by fitting numerical results.
\begin{equation}
\delta_L=0.1441(e^{\theta/56.06}-1)\label{eq1}
\end{equation}
where $\theta$ is contact angle (in degree), and $\delta_L$ is slip length in lattice unit. Our results are also consistent with those of Zhang and Kwok\cite{zhang2004apparent} (see solid circle in FIG.~\ref{fig6} for comparison).
As discussed above, the slip length is the result of interaction between the liquid and the wall surface, and it depends on the properties of the liquid (e.g. $G$) and the wall surface (e.g. $G_{ads}$). So the simulation of slip length should take into account all parameters of the interaction objects. Though it is possible to construct a model of computing slip length in dependence of all interaction parameters, it is a very hard task \cite{harting2006lattice}. In fact, the contact angle is a more comprehensive expression of interactions between liquid and solid surface, and the weaker the solid-fluid interaction is, the larger the contact angle is. So we use contact angle as control parameter to simulate the slip length. The result listed in FIG.~\ref{fig6} shows that this method is feasible, practical and much simpler.

\section{The upper limit of nano-particles' diameter for drag reduction}
\begin{figure}[htbp]
\centering
\subfloat[Original configuration]{
\label{fig7a}
\includegraphics[width=1.6in]{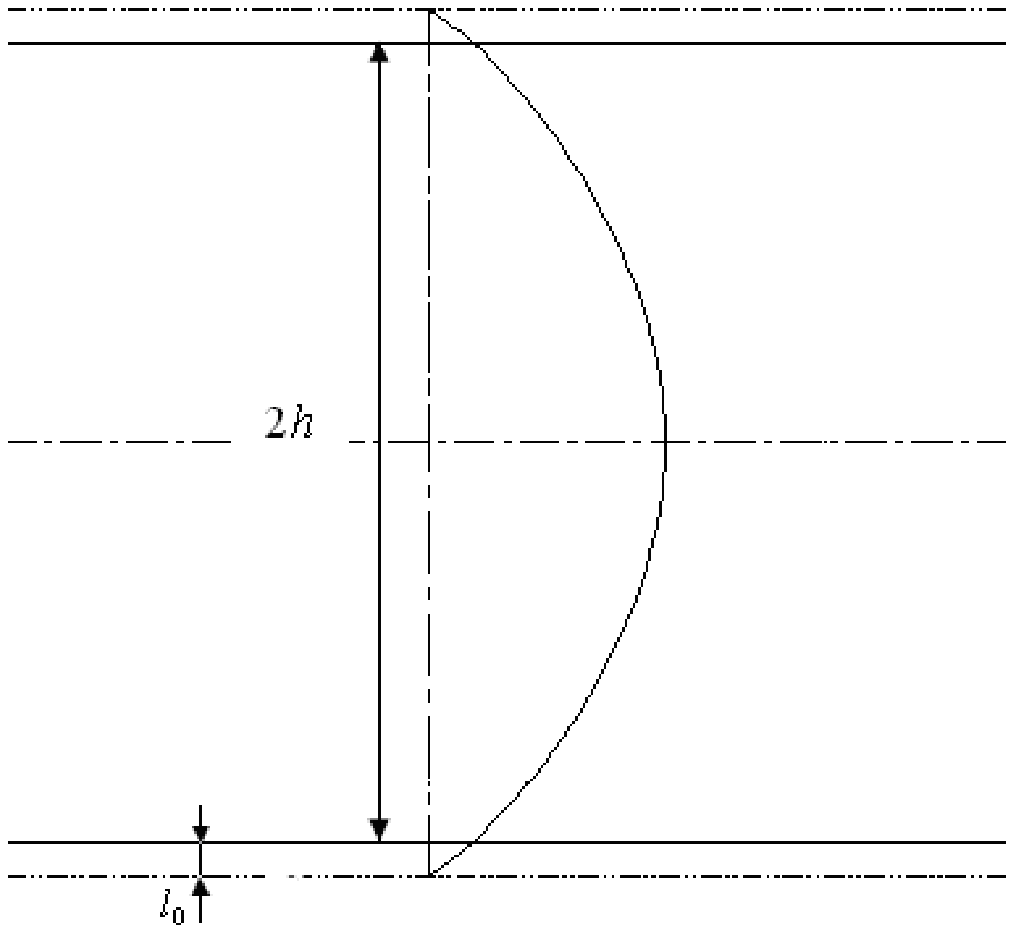}
}
\subfloat[Configuration after HNPs adsorption treatment]{
\label{fig7b}
\includegraphics[width=1.6in]{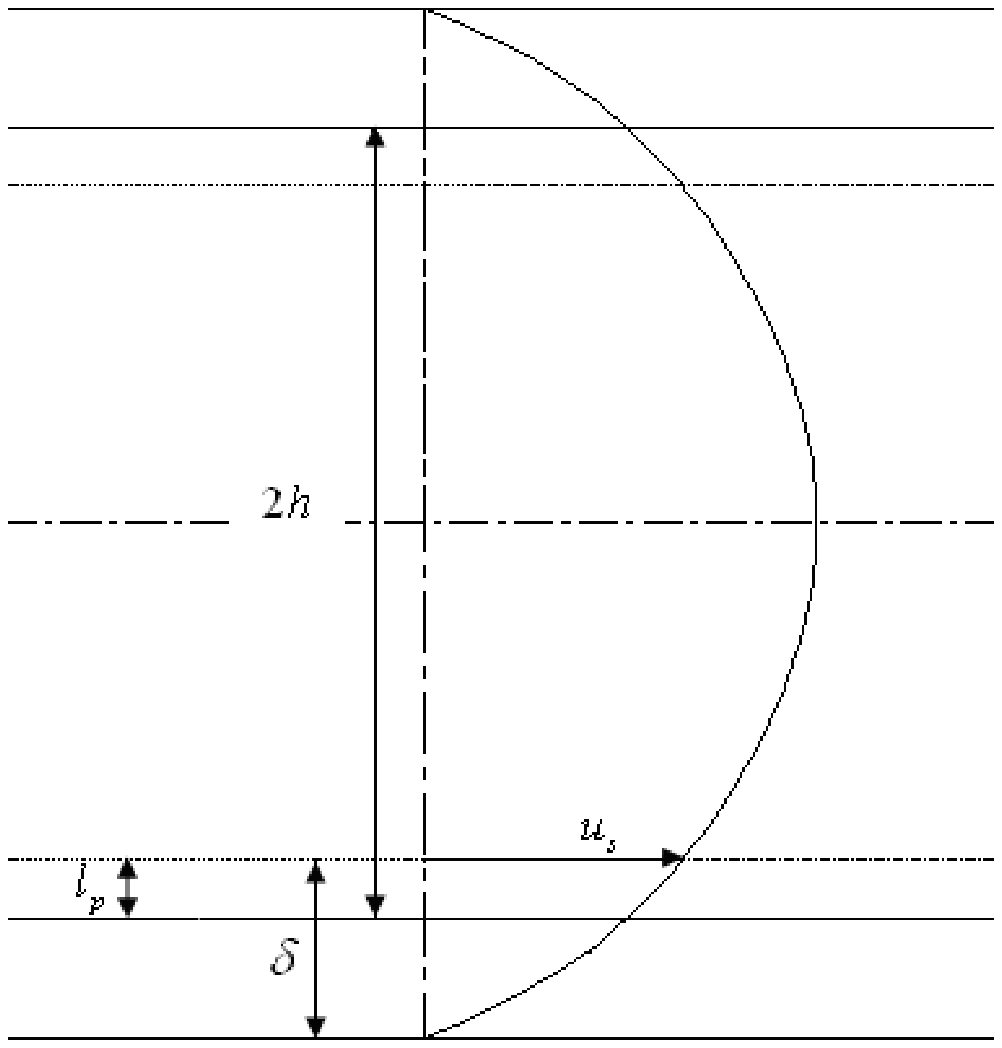}
}
\caption{\label{fig7}The slip model.}
\end{figure}

A practical application of Eq.~(\ref{eq1}) is the determination of the upper limit of nano-particles' size for the purpose of drag reduction. Water flooding has been widely used worldwide as a secondary recovery method, and has been proved an effective method. However, many technical challenges have been encountered during field tests and applications. One of them is the extremely high injection pressure, especially for tight formations. Di, et.al.\cite{qinfeng2009experimental} proposed an effective way of improving water injection and reducing drag of fluid flowing through rock's micro-channels with application of nano-particles adsorption method, and have investigated the drag reduction mechanism experimentally and theoretically. When solution containing hydrophobic nano-particles (HNPs) of $S_iO_2$ is injected into a reservoir, HNPs are adsorbed to the wall of micro-channels to form a strong or super hydrophobic layer, which can lead to a slip boundary condition and thus decrease drag. The slip model is shown in FIG.~\ref{fig7}.

In FIG.~\ref{fig7a}, $2h$ represents the original pore throat diameter, and $l_0$ is slip length brought about by substrate which is usually hydrophilic and has contact angle in the range of 0 to 30 degree. When HNPs with effective diameter of $l_p$ are adsorbed to the wall of the pore throat and formed a single-layer of nano-particles, the pore throat diameter decrease from $2h$ to $2(h-l_p)$, as shown in FIG.~\ref{fig7b}. $u_s$ represents slip velocity in both FIG.~\ref{fig7a} and FIG.~\ref{fig7b}.

In general, $l_p$ does not equal the diameter of nano-particle because of nano-particle's intrinsic contact angle. After the treatment of HNPs adsorption, the wall's wettability is changed from hydrophilic to hydrophobic. The hydrophobicity of the solid surface increases the contact angle significantly and thus yields a much larger slip length $\delta$. The necessary and sufficient condition for drag reduction can be represented as
\begin{equation}
\delta >(l_p+l_0),\label{slip0}
\end{equation}

\begin{figure}[h]
\includegraphics[scale=.7]{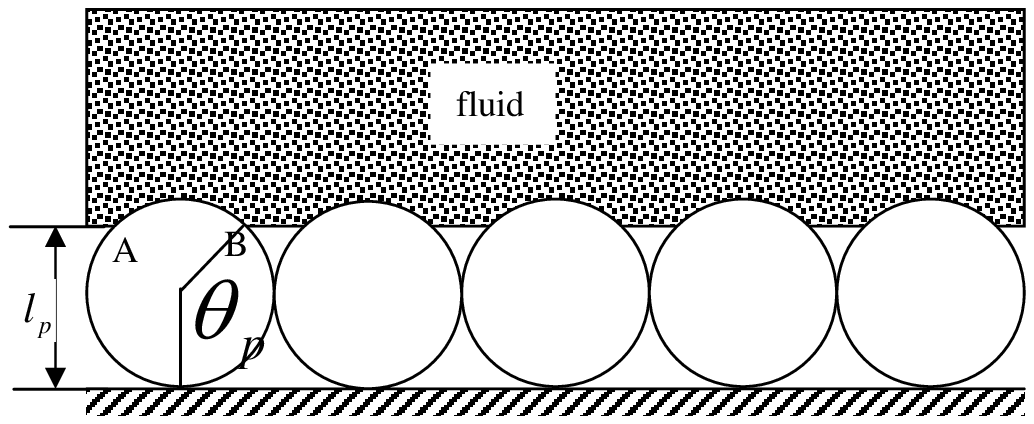}
\caption{\label{fig8}  Schematic of nano-particles adsorbed on the wall.}
\end{figure}

For a 2D micro-channel, we assumed that nano-particles adsorbed on the wall are arranged closely in a single-layer, as shown in FIG.~\ref{fig8}, where $\theta_p$ is the nano-particles' intrinsic contact angle. According to the principle of minimization of the Gibbs energy of a system\cite{marmur2003wetting}, it is easy to determine the location of liquid-gas surface, or the effective diameter of HNPs, $l_p$ , as shown in FIG.~\ref{fig8}, $l_p=d_p/2(1-cos\theta_p)$, $d_p$ is the diameter of a nano-particle. Eq.~(\ref{slip0}) becomes,

\begin{equation}
\delta >\frac{d_p}{2}(1-cos\theta_p)+l_0,\label{slip1}
\end{equation}

The upper limit of nano-particles' diameter for drag reduction can be obtained by rearranging Eq.~(\ref{slip1}),
\begin{equation}
d_{pmax}=\frac{2(\delta-l_0)}{1-cos\theta_p},\label{slip2}
\end{equation}
The slip length in Eq.~(\ref{slip2}), $\delta$ , can be calculated with Eq.~(\ref{eq1}), but the contact angle of a rough surface (as shown in FIG.~\ref{fig8}), $\theta$ , is an unknown parameter. According to the Cassie-Baxter equation\cite{cassie1944wettability}, the contact angle for the heterogeneous wetting regime is given as

\begin{equation}
cos\theta=f_1cos\theta_p-f_2,\label{620}
\end{equation}
where $f_1$ and $f_2$ are the area fraction of the liquid-solid contact and liquid-gas contact, respectively, and can be determined by following equations.

\begin{eqnarray}
f_1=\pi-\theta_p,
\end{eqnarray}
\begin{eqnarray}
f_2=1-sin\theta_p,
\end{eqnarray}

Then Eq.~(\ref{620}) can be expressed as
\begin{equation}
cos\theta=(\pi-\theta_p)cos\theta_p+sin\theta_p-1,\label{623}
\end{equation}

In Eq.~(\ref{623}), the contact angle $\theta$ is independent of the diameter of nano-particle and only depends on nano-particle's intrinsic contact angle, when nano-particles adsorbed on the wall are arranged closely in a single-layer as shown in FIG.~\ref{fig8}. If the intrinsic contact angle $\theta_p$ equals to 120$^\circ$( is usually less than 120$^\circ$\cite{marmur2006underwater} ), the apparent contact angle $\theta$ equals 131.115$^\circ$ from Eq.~(\ref{623}). That is to say, when the nano-particles adsorbed on the wall are as shown in FIG.~\ref{fig8}, the largest apparent contact angle may reach 131.115. Then a slip length of 1.35 lu (in our simulations the units are lattice units as Ref.~\onlinecite{sukop2006lattice}) can be got according to Eq.~(\ref{eq1}). In most cases, the substrate is hydrophilic and usually has contact angle in a range of 0 to 30$^\circ$£¬and the corresponding slip length   is smaller than 0.102 lu. In our example, 1 lu= 20.408nm. When the intrinsic contact angle of nano-particles is 120$^\circ$, the upper limit of the diameter of nano-particles is about 34nm. Thus, the nano-particles adsorbing method is effective only when the diameter of nano particles is smaller than 34 nm.

\begin{figure}[h]
\includegraphics[scale=.3]{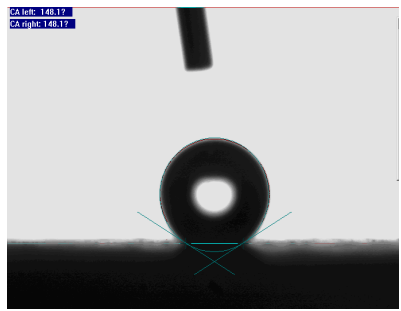}
\caption{\label{fig9}  The contact angle of a core sample after adsorbing nano-particles is 148.1 degree.}
\end{figure}

We should note that both Eqs.~(\ref{slip2}) and~(\ref{623}) are derived under the assumption that all nano-particles have same diameter and are well arranged as shown in FIG.~\ref{fig8}. However, it is not the case for practical applications. In fact, the nano-particles have a statistical range of diameter, and they cannot be so well arranged and adsorbed on the wall. The wall surface is much more rough. So the above upper limit is a theoretical result and will be different from the practical situations. But according to Lauga, E., et al\cite{lauga2006microfluidics}, a more rough surface may lead to a greater contact angle. This conclusion has also been confirmed by our experiments, where the largest contact angle is 148.1$^\circ$, as shown in FIG.~\ref{fig9}. Therefore, Eq.~(\ref{slip2}) yields a more conservative result. Though it is not an exact solution, the upper limit estimated by Eq.~(\ref{slip2}) is still useful for determination the suitable size of nano-particles for a given reservoir.
\section{Conclusions}
(1) Both contact angle of a fluid on a solid surface and slip length of a fluid flowing over a solid surface can be easily simulated with the Shan-Chen LB model.

(2) Both the non-ideal interaction between fluid particles (controlled by a parameter of $G$) and the interaction between a fluid and a solid (controlled by another parameter of $G_{ads}$) have significant impact on contact angle and slip length. However, they have little influence on the relationship between contact angle and slip length.

(3) There is an upper limit on the diameter of nano-particles when applying nano-particle adsorption method to reduce drag. The upper limit can be estimated by applying the relationship between contact angle and slip length given the properties of nano-particles and geometrical parameters of a micro-channel.

\begin{acknowledgments}
This research is supported partly by the National Science Funding of China (50674065£¬50874071), the Chinese National Programs for High Technology Research and Development (2008AA06Z201), the Key Program of Science and Technology Commission of Shanghai Municipality(071605102), Shanghai Program for Innovative Research Team in Universities, the Innovation Fund Project for Graduate Student of Shanghai University (SHUCX101076), the Program for Changjiang Scholars and Innovative Research Team in University (IRT0844).
\end{acknowledgments}

\nocite{*}
\bibliographystyle{unsrt}
\bibliography{apssamp}

\end{document}